\documentclass[american,graybox]{svmult}

\usepackage{type1cm}
\usepackage{cite}
\usepackage{color}
\usepackage{makeidx}
\usepackage{graphicx}
\usepackage{multicol}
\usepackage[bottom]{footmisc}
\usepackage{qtree}
\usepackage{float}
\usepackage{url}
\usepackage{stmaryrd}
\usepackage{amsmath,amssymb}

\usepackage{mathtools}
\DeclareMathAlphabet{\mathup}{OT1}{\familydefault}{m}{n}
\usepackage{scalerel}
\newcommand{\substitute}[3]{\mathup{S}^{\ThisStyle{\raisebox{-.5ex}{$\SavedStyle #1$}}}_{#2}#3\,|}

\newcommand{\Fig}[1]{Fig. \ref{#1}}
\newcommand{\Sec}[1]{Sect. \ref{#1}}

\newcommand{\Tab}[1]{Tab. \ref{#1}}
\newcommand{\TT}[1]{\texttt{#1}}
\newcommand{\MT}[1]{\mathtt{#1}}
\newcommand{\seq}[1]{\mathbf{#1}}
\newcommand{\lex}{\TT{:\!:}}        
\newcommand{\der}{\TT{:}}           

\makeindex

\begin{document}

\title*{Reinforcement learning of minimalist grammars}
\author{
    Peter beim Graben,  Ronald R\"{o}mer, Werner Meyer, Markus Huber, and Matthias Wolff
}
\authorrunning{beim Graben et al.}
\institute{
    Peter beim Graben \at
    Fraunhofer Institute for Ceramic Technologies and Systems IKTS \\
    Project Group ``Cognitive Material Diagnostics'' \\
    Brandenburgische Technische Universit\"at (BTU) Cottbus--Senftenberg\\
    Platz der Deutschen Einheit 1 \\
    D -- 03046 Cottbus,
    \email{peter.beimgraben@b-tu.de} \and
    Ronald R\"{o}mer
    \and
    Werner Meyer
    \and
    Markus Huber
    \and
    Matthias Wolff \at
    Department of Communication Engineering \\
    Brandenburgische Technische Universit\"at (BTU) Cottbus--Senftenberg\\
    Platz der Deutschen Einheit 1 \\
    D -- 03046 Cottbus
}

\maketitle

\abstract*{Speech-controlled user interfaces facilitate the operation of devices and household functions to laymen. State-of-the-art language techno\-logy scans the acoustically analyzed speech signal for relevant keywords that are subsequently inserted into semantic slots to interpret the user's intent. In order to develop proper cognitive information and communication technologies, simple slot-filling should be replaced by utterance meaning transducers (UMT) that are based on semantic parsers and a \emph{mental lexicon}, comprising syntactic, phonetic and semantic features of the language under consideration. This lexicon must be acquired by a cognitive agent during interaction with its users. We outline a reinforcement learning algorithm for the acquisition of syntax and semantics of English utterances, based on minimalist grammar (MG), a recent computational implementation of generative linguistics. English declarative sentences  are presented to the agent by a teacher in form of utterance meaning pairs (UMP) where the meanings are encoded as formulas of predicate logic. Since MG codifies universal linguistic competence through inference rules, thereby separating innate linguistic knowledge from the contingently acquired lexicon, our approach unifies generative grammar and reinforcement learning, hence potentially resolving the still pending Chomsky-Skinner controversy.}

\abstract{Speech-controlled user interfaces facilitate the operation of devices and household functions to laymen. State-of-the-art language techno\-logy scans the acoustically analyzed speech signal for relevant keywords that are subsequently inserted into semantic slots to interpret the user's intent. In order to develop proper cognitive information and communication technologies, simple slot-filling should be replaced by utterance meaning transducers (UMT) that are based on semantic parsers and a \emph{mental lexicon}, comprising syntactic, phonetic and semantic features of the language under consideration. This lexicon must be acquired by a cognitive agent during interaction with its users. We outline a reinforcement learning algorithm for the acquisition of syntax and semantics of English utterances, based on minimalist grammar (MG), a recent computational implementation of generative linguistics. English declarative sentences  are presented to the agent by a teacher in form of utterance meaning pairs (UMP) where the meanings are encoded as formulas of predicate logic. Since MG codifies universal linguistic competence through inference rules, thereby separating innate linguistic knowledge from the contingently acquired lexicon, our approach unifies generative grammar and reinforcement learning, hence potentially resolving the still pending Chomsky-Skinner controversy.}



\section{Introduction}
\label{sec:intro}

Speech-controlled user interfaces \cite{Young10}\index{user interface | speech-controlled user interface} such as Amazon's \emph{Alexa}, Apple's \emph{Siri} or \emph{Cortana} by Microsoft substantially facilitate the operation of devices and household functions to laymen. Instead of using keyboard and display as input-output interfaces, the operator pronounces requests or instructions to the device and listens to its responses. One important future development will be Smart Home \index{smart home} and Ambient Assisted Living \index{ambient assisted living} applications in the health sciences \cite{MartinWeibel18}.

State-of-the-art language technology \index{language technology} scans the acoustically analyzed speech signal for relevant keywords that are subsequently inserted into semantic frames \cite{Minsky74, Fillmore88} to interpret the user's intent. This \emph{slot-filling} \index{slot-filling} procedure \cite{Allen03, TurHakkaniEA11, MesnilDauphinEA15} is based on large language corpora that are evaluated by standard machine learning \index{machine learning} methods, such as conditional random fields \cite{TurHakkaniEA11} or by deep neural networks \cite{MesnilDauphinEA15}, for instance. The necessity to overcome traditional slot-filling \index{slot-filling} by proper cognitive information and communication technologies \index{cognitive information and communication technology} \cite{BaranyiCsapoSallai15} has already been emphasized by Allan \cite{Allen17}. His research group trains semantic parsers \index{semantic parser} from large language data bases such as WordNet or VerbNet that are constrained by hand-crafted expert knowledge and semantic ontologies \cite{Allen03, Allen14, AllenBahkshandehEA18, PereraAllenEA18b}.

One particular demand on cognitive user interfaces \index{user interface} are the processing and understanding of declarative or imperative sentences. Consider, e.g., a speech-controlled heating device, a \emph{cognitive heating} \cite{DuckhornHuberEA17, KlimczakWolffLindemannEA14, TschopeEA18, WolffMeyerRomer15, WolffRomerWirsching15}, with the operator's utterance ``I am cold!'' This declarative sentence must firstly be analyzed syntactically to assigning ``I'' to the subject position and attributing ``am cold'' to the predicate. Secondly, a semantic analysis interprets ``I'' as the speaker and ``am cold'' as a particular subjective state. From this representation of the speaker's state, the system must compute logical inferences and then respond accordingly, by increasing the room temperature and probably by giving a linguistic feedback signal ``I increase the temperature to 22 degrees''. Technically, this could be achieved using feature-value relations (FVR) \cite{WolffMeyerRomer15} as semantic representations and modified Markov-decision processes (MDP) for behavior control \cite{TschopeEA18, WolffMeyerRomer15, WolffRomerWirsching15}.

Recent research in computational linguistics \index{computational linguistics} has demonstrated that quite different grammar formalisms, such as tree-adjoining grammar \cite{JoshiLevyTakahashi75}\index{grammar| tree-adjoining}, multiple context-free grammar \cite{SekiEA91}\index{grammar| multiple context-free}, range concatenation grammar \cite{Boullier05}\index{grammar| range concatenation}, and minimalist grammar \cite{Stabler97, StablerKeenan03}\index{grammar| minimalist grammar} converge toward universal description models \cite{KuhlmannKollerSatta15, JoshiVijayWeir90, Michaelis01a, Stabler11a}. Minimalist grammar (MG) \index{grammar| minimalist grammar} has been developed by Stabler \cite{Stabler97} to mathematically codify Chomsky's \emph{Minimalist Program} \cite{Chomsky95} in the generative grammar \index{grammar | generative} framework. A minimalist grammar \index{grammar| minimalist grammar} consists of a mental lexicon storing linguistic signs \index{linguistic sign} as arrays of syntactic, phonetic and semantic features, on the one hand, and of two structure-building functions, called ``merge'' and ``move'', on the other hand. Syntactic features in the lexicon are, e.g., the linguistic base types noun ($\TT{n}$), verb  ($\TT{v}$), adjective ($\TT{a}$), determiner ($\TT{d}$), inflection (i.e. tense $\TT{t}$), or, preposition ($\TT{p}$). These are syntactic heads selecting other categories either as complements or as adjuncts. The structure generation is controlled by selector categories that are ``merged'' together with their selected counterparts. Moreover, one distinguishes between licensors and licensees, triggering the movement of maximal projections. An MG does not comprise any phrase structure rules; all syntactic information is encoded in the feature array of the mental lexicon. Furthermore, syntax and compositional semantics can be combined via the lambda calculus \cite{Niyogi01, Kobele09},\index{lambda calculus} while MG parsing \index{parsing} can be straightforwardly implemented through bottom-up \cite{Harkema01}\index{parsing | bottom-up}, top-down \cite{Harkema01, Mainguy10, Stabler11b},\index{parsing | top-down} and in the meantime also by left-corner \index{parsing | left-corner} automata \cite{StanojevicStabler18}.

One important property of MG is their effective learnability in the sense of Gold's formal learning theory \cite{Gold67}\index{learning theory}. Specifically, MG can be acquired by positive examples \cite{BonatoRetore01, KobeleCollierEA02, StablerEA03} from linguistic dependence graphs \cite{BostonHaleKuhlmann10, Nivre03, KleinManning04}, which is consistent with psycholinguistic findings on early-child language acquisition \cite{Ellis06, Diessel13, Gee94, Pinker95, Tomasello06}. However, learning through positive examples only, could easily lead to overgeneralization. According to Pinker \cite{Pinker95} this could substantially be avoided through reinforcement learning \cite{Skinner15, SuttonBarto18}\index{reinforcement learning}. Although there is only little psycholinguistic evidence for reinforcement learning in human language acquisition \cite{Moerk83, SundbergEA96},\index{language acquisition} we outline a machine learning \index{machine learning} algorithm for the acquisition of an MG mental lexicon \cite{GrabenEA20} of the syntax and semantics for English declarative sentences through reinforcement learning in this chapter. Instead of looking at pure syntactic dependencies as in \cite{BonatoRetore01, KobeleCollierEA02, StablerEA03}, our approach directly uses their underlying semantic dependencies for the simultaneous segmentation of syntax and semantics.


\section{Minimalist Grammar}
\label{sec:mg}

Our language acquisition \index{language acquisition} approach for minimalist grammar \index{grammar| minimalist grammar} combines methods from computational linguistics, \index{computational linguistics} formal logic, and abstract algebra. Starting point of our algorithm are \emph{utterance meaning pairs} (UMP) \index{utterance meaning pair (UMP)} \cite{WirschingLorenz13, GrabenEA20}.
\begin{equation}\label{eq:ump}
    u = \langle e, \sigma \rangle \:,
\end{equation}
where $e \in E$ is the spoken or written utterance, given as the \emph{exponent} \index{exponent} of a linguistic sign \index{linguistic sign} \cite{Kracht03}. Technically, exponents are strings taken from the Kleene hull of some finite alphabet, $A$, i.e. $E = A^*$. The sign's \emph{semantics} $\sigma \in \Sigma$ is a logical term, expressed by means of predicate logic \index{predicate logic} and the (untyped) lambda calculus \cite{Church36}\index{lambda calculus}.


\subsection{Semantics}
\label{sec:sem}

As an example, we consider the simple UMP \index{utterance meaning pair (UMP)}
\begin{equation}\label{eq:ump1}
    u = \langle \TT{the mouse eats cheese}, \MT{eat(cheese)(mouse)} \rangle
\end{equation}
in the sequel. We use typewriter font throughout the chapter to emphasize that utterances are regarded plainly as symbolic tokens without any intended meaning in the first place. This applies even to the ``semantic'' representation in terms of first order predicate logic \index{predicate logic} where we use the Sch\"onfinkel-Curry \cite{Lohnstein11, Schonfinkel24} notation here. Therefore, the expression above $\MT{eat(cheese)(mouse)}$ indicates that $\MT{eat}$ is a binary predicate, fetching first its direct object $\MT{cheese}$ to form a unary predicate, $\MT{eat(cheese)}$, that then takes its subject $\MT{mouse}$ in the second step to build the proposition of the utterance \eqref{eq:ump1}.

Following Kracht \cite{Kracht03}, we regard a linguistic sign \index{linguistic sign} as an ordered triple
\begin{equation}\label{eq:sign}
    z = \langle e , t , \sigma \rangle
\end{equation}
with the same exponent \index{exponent} $e \in E$ and semantics $\sigma \in \Sigma$ as in the UMP \index{utterance meaning pair (UMP)} \eqref{eq:ump}. In addition, $t \in T$ is a syntactic \emph{type} that we encode by means of minimalist grammar (MG) \index{grammar| minimalist grammar} in its chain representation \cite{StablerKeenan03}. The type controls the generation of syntactic structure and hence the order of lambda application, \index{lambda application} analogously to the typed lambda calculus \cite{Church40}\index{lambda calculus} in Montague semantics \cite{Lohnstein11}\index{semantics | Montague}. In order to avoid redundancy, we use the plain (untyped) lambda calculus in the following \cite{Church36}.

Lambda calculus is a mathematical formalism developed by Church in the 1930s ``to model the mathematical notion of substitution of values for bound variables'' according to~\cite{Wegner2003}. Although the original application was in the area of computability (cf.~\cite{Church36}) the substitution of parts of a term with other terms is often the central notion when lambda calculus \index{lambda calculus} is used. This is also true in our case and we have to clarify the concepts first. Namely \emph{variable}, \emph{bound} and \emph{free}, \emph{term}, and \emph{substitution}.

To be applicable to any universe of discourse one prerequisite of lambda calculus \index{lambda calculus} is that variables are elements from ``an enumerably infinite set of symbols''~\cite{Church36}. However, for usage in a specific domain, a finite set is sufficient. Since we aim at terms from first order predicate logic, treating them with the operations from lambda calculus, \index{lambda calculus} all their predicates $P$ and individuals $I$ need to be in the set of variables. Additionally, we will use the symbols $I_{V}\coloneqq\{\MT{x},\MT{y},\ldots\}$ as variables for individuals and $T_{V}\coloneqq\{\MT{P},\MT{Q},\ldots\}$ as variables for (parts of) logical terms. The set $V\coloneqq P\cup I\cup I_{V}\cup T_{V}$ is thus used as the set of variables. Note, that the distinction made by $I_{V}$ and $T_{V}$ is not on the level of lambda calculus \index{lambda calculus} but rather a visual clue to the reader.

The term algebra of lambda calculus \index{lambda calculus} is inductively defined as follows. \emph{i})~Every variable $v\in V$ is a term and $v$ is a \emph{free} variable in the term $v$; specifically, also every well-formed formula of predicate logic is a term. \emph{ii})~Given a term $T$ and a variable $v\in V$ which is free in $T$, the expression $\MT{\lambda} v.T$ is also a term and the variable $v$ is now \emph{bound} in $\MT{\lambda} v.T$. Every other variable in $T$ different from $v$ is free (bound) in $\MT{\lambda} v.T$ if it is free (bound) in $T$. \emph{iii})~Given two terms $T$ and $U$, the expression $T(U)$ is also a term and every variable which is free (bound) in $T$ or $U$ is free (bound) in $T(U)$. Such a term is often referred to as \emph{operator-operand combination}~\cite{Wegner2003} or \emph{functional application}~\cite{Lohnstein11}. For disambiguation we also allow parentheses around terms. The introduced syntax differs from the original one where additionally braces and brackets are used to mark the different types of terms (cf.~\cite{Church36}). Sometimes, $T(U)$ is also written as $(TU)$ and the dot between $\MT{\lambda}$ and the variable is left out (cf.~\cite{Wegner2003}).

For a given variable $v\in V$ and two terms $T$ and $U$ the operation of \emph{substitution} is $T[v \leftarrow U]$ (originally written as $\substitute{v}{U}{T}$ in~\cite{Church36} and sometimes without the right bar, i.\,e.\ as in~\cite{Wegner2003}) and stands for the result of substituting $U$ for all instances of $v$ in $T$.

Church defined three conversions based on substitution.
\begin{itemize}
\item \emph{Renaming} bound variables by replacing any part $\MT{\lambda} v.T$ of a term by $\MT{\lambda} w.T[v \leftarrow w]$ when the variable $w$ does not occur in the term $T$.
\item \emph{Lambda application} \index{lambda application} by replacing any part $\MT{\lambda} v.T(U)$ of a term by $T[v \leftarrow U]$, when the bound variables in $T$ are distinct both from $v$ and the free variables in $U$.
\item \emph{Lambda abstraction} \index{lambda abstraction} by replacing any part $T[v \leftarrow U]$ of a term by $\MT{\lambda} v.T(U)$, when the bound variables in $T$ are distinct both from $v$ and the free variables in $U$.
\end{itemize}
The first conversion simply states that names of bound variables have no particular meaning on their own. The second and third conversions are of special interest to our aims. Lambda application \index{lambda application} allows the composition of logical terms out of predicates, individuals and other logical terms while lambda abstraction \index{lambda abstraction} allows the creation of templates of logical terms.

Applied to our example \eqref{eq:ump1}, we have the sign
\begin{equation}\label{eq:sign1}
    z = \langle \TT{the mouse eats cheese}, \der \TT{c}, \MT{eat(cheese)(mouse)} \rangle
\end{equation}
where the now appearing MG type $\der \TT{c}$ indicates that the sign is complex (not lexical) and a complementizer phrase of type $\TT{c}$. Its compositional semantics \cite{Lohnstein11} can be described by the terms $\MT{\lambda P.\lambda Q.P(Q)}$ and $\MT{\lambda P.\lambda Q.Q(P)}$, the predicate $\MT{eat}$ and the individuals $\MT{cheese}$ and $\MT{mouse}$. Consider the term $\MT{\lambda P.\lambda Q.P(Q)(eat)(cheese)}$. This is converted by two successive lambda applications \index{lambda application} via $\MT{\lambda Q.eat(Q)(cheese)}$ into the logical term $\MT{eat(cheese)}$. It is also possible to rearrange parts of the term in a different way. Consider now the term $\MT{\lambda P.\lambda Q.Q(P)}$, the logical term $\MT{eat(cheese)}$ and the individual $\MT{mouse}$. Then the term $\MT{\lambda P.\lambda Q.Q(P)(mouse)(eat(cheese))}$ is converted by two successive lambda applications into the logical term $\MT{eat(cheese)(mouse)}$. Thus, logical terms can be composed through lambda application.

Moreover, given the logical term $\MT{eat(cheese)(mouse)}$ two successive lambda abstractions \index{lambda abstraction} yield the term $\MT{\lambda x.\lambda y.eat(x)(y)}$, leaving out the operand parts. In that way, templates of logical terms are created where different individuals can be inserted for term evaluation. Both processes are crucial for our utterance-meaning transducer \index{utterance-meaning transducer (UMT)} and machine language acquisition \index{language acquisition} algorithms below.


\subsection{Syntax}
\label{sec:syx}

An MG consists of a data base, the mental lexicon, containing signs as arrays of syntactic, phonetic and semantic \emph{features}, and of two structure-generating functions, called ``merge'' and ``move''. Syntactic features are the \emph{basic types} $b \in B$ from a finite set $B$, with $b = \TT{n}, \TT{v}, \TT{a}, \TT{d}$, etc, together with a set of their respective \emph{selectors} $S = \{ \TT{=}b | b \in B \}$ that are unified by the ``merge'' operation. Moreover, one distinguishes between a set of licensers $L_+ = \{ \TT{+}l | l \in L \}$ and another set of their corresponding licensees $L_- = \{ \TT{-}l | l \in L \}$ triggering the ``move'' operation. $L$ is another finite set of movement identifiers. $F = B \cup S \cup L_+ \cup L_-$ is called the feature set. Finally, one has a two-element set $C = \{ \lex, \der \}$ of categories,\footnote{
    Departing from the convention in the literature \cite{StablerKeenan03}, we call the elements of $C$ \emph{categories} due to a tentative interpretation in terms of indexed grammars \cite{Aho68, Staudacher93}\index{grammar | indexed}. We shall address this interesting issue in subsequent research.
}
where ``$\lex$'' indicates \emph{simple}, \emph{lexical} categories while ``$\der$'' denotes \emph{complex}, \emph{derived} categories. The ordering of syntactic features as they appear originally in the lexicon is prescribed as regular expressions, i.e. $T = C (S \cup L_+)^* B L_-^*$ is the set of syntactic lexical types \cite{Stabler97, StablerKeenan03}. The set of linguistic signs \index{linguistic sign} is then given as $Z = E \times T \times \Sigma$ \cite{Kracht03}.

Let $e_1, e_2 \in E$ be exponents, \index{exponent} $\sigma_1, \sigma_2 \in \Sigma$ semantic terms in the lambda calculus, \index{lambda calculus} $f \in B \cup L$ one feature identifier, $\seq{t}, \seq{t}_1, \seq{t}_2 \in F^+$ feature strings compatible with the regular types in $T$, $\cdot \in C$ and $\seq{z}, \seq{z}_1, \seq{z}_2 \in Z^*$ sequences of signs, then $\langle e_1 , \TT{:\!:} \TT{=}f \seq{t}_1, \sigma_1 \rangle$ and $\langle e_2 , \TT{:} f , \sigma_2 \rangle$ form signs in the sense of \eqref{eq:sign}. A sequence of signs is called a \emph{minimalist expression}, and the first sign of an expression is called its \emph{head}, controlling the structure building through ``merge'' and ``move'' as follows.

The MG function ``merge'' is defined through inference schemata
\begin{eqnarray}
&&\dfrac{
    \langle e_1, \TT{:\!:=}f \seq{t}, \sigma_1 \rangle \quad
    \langle e_2, \cdot f, \sigma_2 \rangle \seq{z}
  }{
    \langle e_1e_2, \TT{:} \seq{t}, \sigma_1\sigma_2 \rangle \seq{z}
  }
  \,\text{merge-1} \:, \label{merge-1} \\
&&\dfrac{
    \langle e_1, \TT{:=}f \seq{t}, \sigma_1 \rangle \seq{z}_1 \quad
    \langle e_2, \cdot f, \sigma_2 \rangle \seq{z}_2
  }{
    \langle e_2e_1, \TT{:} \seq{t}, \sigma_1 \sigma_2 \rangle \seq{z}_1 \seq{z}_2
  }
  \,\text{merge-2} \:, \label{merge-2} \\
&&\dfrac{
    \langle e_1,\cdot \TT{=}f \seq{t}_1, \sigma_1 \rangle \seq{z}_1 \quad
    \langle e_2, \cdot f \seq{t}_2, \sigma_2 \rangle \seq{z}_2
  }{
    \langle e_1, \TT{:} \seq{t}_1, \sigma_1 \rangle \seq{z}_1
    \langle e_2, \TT{:} \seq{t}_2, \sigma_2 \rangle \seq{z}_2
  }
  \,\text{merge-3} \:, \label{merge-3}
\end{eqnarray}
Correspondingly, ``move'' is given through
\begin{eqnarray}
&&\dfrac{
    \langle e_1, \TT{:+}f \seq{t}, \sigma_1 \rangle \seq{z}_1
    \langle e_2, \TT{:-}f, \sigma_2 \rangle \seq{z}_2
  }{
    \langle e_2e_1, \TT{:} \seq{t}, \sigma_1\sigma_2 \rangle \seq{z}_1 \seq{z}_2
  }
  \,\text{move-1} \:, \label{move-1} \\
&&\dfrac{
    \langle e_1, \TT{:+}f \seq{t}_1, \sigma_1 \rangle \seq{z}_1
    \langle e_2, \TT{:-}f \seq{t}_2, \sigma_2 \rangle \seq{z}_2
  }{
    \langle e_1, \TT{:} \seq{t}_1, \sigma_1 \rangle \seq{z}_1
    \langle e_2, \TT{:} \seq{t}_2, \sigma_2 \rangle \seq{z}_2
  }
  \,\text{move-2} \:. \label{move-2}
\end{eqnarray}
where only one sign with licensee $\TT{-}f$ may appear in the expression licensed by $\TT{+}f$ in the head. This so-called \emph{shortest movement constraint} (SMC) guarantees syntactic locality demands \cite{Stabler97, StablerKeenan03}.

A minimalist derivation terminates when all syntactic features besides one distinguished \emph{start symbol}, which is $\TT{c}$ in our case, have been consumed. We conventionally use complementizer phrase $\TT{c}$ as the start symbol.

For illustrating the rules (\ref{merge-1} -- \ref{move-2}) and their applicability, let us stick with the example UMP \index{utterance meaning pair (UMP)} \eqref{eq:ump1}. Its syntactic analysis in terms of generative grammar \cite{Haegeman94}\index{grammar | generative} yields the (simplified) phrase structure tree in \Fig{fig:gb}(a).\footnote{
    For the sake of simplicity we refrain from presenting full-fledged X-bar hierarchies \cite{Haegeman94}.
}

\begin{figure}[H]
(a)
\Tree [.\TT{CP}
    [.\TT{DP}
        [.\TT{D} \TT{the} ].\TT{D}
        [.\TT{N} \TT{mouse} ].\TT{N}
    ].\TT{DP}
    [.\TT{IP}
        [.\TT{I} $\TT{eat}_i$ ].\TT{I}
        [.\TT{I}\1
            [.\TT{I} \TT{-s} ].\TT{I}
            [.\TT{VP}
                [.\TT{V} $\epsilon_i$ ].\TT{V}
                    [.\TT{N} \TT{cheese} ].\TT{N}
            ].\TT{VP}
        ].\TT{I}\1
    ].\TT{IP}
].\TT{CP}
\hfill
(b)
\Tree [.{$\MT{eat(cheese)(mouse)}$}
    [.{$\MT{\lambda y.eat(cheese)(y)}$}
        {$\MT{\lambda x. \lambda y. eat(x)(y)}$} !\qsetw{3cm} {$\MT{cheese}$}
    ]
    {$\MT{mouse}$}
]
\caption{\label{fig:gb} Generative grammar \index{grammar | generative} analysis of example UMP \index{utterance meaning pair (UMP)} \eqref{eq:ump1}. (a) Syntactic phrase structure tree. (b) Semantic tree from lambda calculus. \index{lambda calculus}}
\end{figure}

The syntactic categories in \Fig{fig:gb}(a) are the \emph{maximal projections} \TT{CP} (complementizer phrase), \TT{IP} (inflection phrase), \TT{VP} (verbal phrase), and \TT{DP} (determiner phrase). Furthermore, there are the intermediary node $\TT{I}'$ and the \emph{heads} \TT{I} (inflection), \TT{D} (determiner), \TT{V} (verb), and \TT{N} (noun), corresponding to \TT{t}, \TT{d}, \TT{v}, and \TT{n} in MG, respectively. Note that inflection is lexically realized only by the present tense suffix \TT{-s}. Moreover, the verb \TT{eat} has been moved out of its base-generated position leaving the empty string $\epsilon$ there. Movement is indicated by co-indexing with $i$.

Correspondingly, we present a simple semantic analysis in \Fig{fig:gb}(b) using the notation from \Sec{sec:sem} together with the lambda calculus \index{lambda calculus} of the binary predicate in its Sch\"onfinkel-Curry representation \cite{Lohnstein11, Schonfinkel24}.

Guided by the linguistic analyses in \Fig{fig:gb}, an expert could construe a minimalist lexicon as given in \Tab{tab:mg0} by hand \cite{StablerKeenan03}.

\begin{table}[H]
\caption{\label{tab:mg0} Minimalist lexicon for example grammar \Fig{fig:gb}.}
\[
\begin{array}{ll}
    \langle \TT{mouse}, \lex \TT{n}, \MT{mouse} \rangle &
    \langle \TT{cheese}, \lex \TT{n -k}, \MT{cheese} \rangle \\
    \langle \TT{the}, \lex \TT{=n d -k}, \epsilon \rangle &
    \langle \TT{eat}, \lex \TT{=n v -f}, \MT{\lambda x. \lambda y. eat(x)(y)} \rangle \\
    \langle \TT{-s}, \lex \TT{=pred +f +k t}, \epsilon \rangle &
    \langle \epsilon, \lex \TT{=v +k =d pred}, \MT{\lambda P. \lambda Q. Q(P)} \rangle \\
    \langle \epsilon, \lex \TT{=t c}, \epsilon \rangle
\end{array}
\]
\end{table}

We adopt a shallow semantic model, where the universe of discourse only contains two individuals, the mouse and a piece of cheese.\footnote{
Moreover, we abstract our analysis from temporal and numeral semantics and also from the intricacies of the semantics of noun phrases in the present exposition.
}
Then, the lexicon \Tab{tab:mg0} is interpreted as follows. Since all entries are contained in the MG lexicon, they are of category ``\lex''. There are two nouns (\TT{n}), \TT{mouse} and \TT{cheese} with their respective semantics as individual constants, $\MT{mouse}$ and $\MT{cheese}$. In contrast to \TT{mouse}, the latter possesses a licensee \TT{-k} for case marking. The same holds for the determiner \TT{the} selecting a noun (\TT{=n}) as its complement to form a determiner phrase \TT{d} which also requires case assignment (\TT{-k}) afterwards. The verb (\TT{v}) \TT{eat} selects a noun as a complement and has to be moved for inflection \TT{-f}. Its compositional semantics is given by the binary predicate $\MT{eat(x)(y)}$ whose argument variables are bounded by two lambda expressions $\MT{\lambda x. \lambda y}$. Moreover, we have an inflection suffix \TT{-s} for present tense in third person singular, taking a predicate (\TT{pred}) as complement, then triggering firstly inflection movement \TT{+f} and secondly case assignment \TT{+k}, whose type is tense (\TT{t}). Finally, there are two entries that are phonetically not realized. The first one selects a verbal phrase \TT{=v} and assigns case \TT{+k} afterwards; then, it selects a determiner phrase \TT{=d} as subject and has its own type predicate \TT{pred}; additionally, we prescribe an intertwiner of two abstract lambda expressions $\MT{Q, P}$ as its semantics. The last entry provides a simple type conversion from tense \TT{t} to complementizer \TT{c} in order to arrive at a well-formed sentence with start symbol \TT{c}.

Using the lexicon \Tab{tab:mg0}, the sign \eqref{eq:sign1} is obtained by the minimalist derivation \eqref{eq:mgder1}.

{\scriptsize
\begin{subequations}\label{eq:mgder1}
\renewcommand{\theequation}{\theparentequation{}-{}\arabic{equation}}
\begin{align}
&\dfrac{\langle \TT{the}, \lex \TT{=n d -k}, \epsilon \rangle
    \qquad
    \langle \TT{mouse}, \lex \TT{n}, \MT{mouse} \rangle
    }
    {\langle \TT{the mouse}, \der \TT{d -k}, \MT{mouse}\rangle
    } \: \text{merge-1} \label{eq:mgder1-1} \\
\nonumber \\
&\dfrac{\langle \TT{eat}, \lex \TT{=n v -f}, \MT{\lambda x. \lambda y. eat(x)(y)} \rangle
    \qquad
    \langle \TT{cheese}, \lex \TT{n -k}, \MT{cheese} \rangle
    }
    {\langle \TT{eat}, \der \TT{v -f}, \MT{\lambda x. \lambda y. eat(x)(y)} \rangle
    \langle \TT{cheese}, \der \TT{-k}, \MT{cheese} \rangle
    } \: \text{merge-3} \label{eq:mgder1-2} \\
&\dfrac{\langle \epsilon, \lex \TT{=v +k =d pred}, \MT{\lambda P. \lambda Q. Q(P)} \rangle
    \qquad
    \langle \TT{eat}, \der \TT{v -f}, \MT{\lambda x. \lambda y. eat(x)(y)} \rangle
    \langle \TT{cheese}, \der \TT{-k}, \MT{cheese} \rangle
    }
    {\langle \epsilon, \der \TT{+k =d pred}, \MT{\lambda P. \lambda Q. Q(P)} \rangle
    \langle \TT{eat}, \der \TT{-f}, \MT{\lambda x. \lambda y. eat(x)(y)} \rangle
    \langle \TT{cheese}, \der \TT{-k}, \MT{cheese} \rangle
    } \: \text{merge-3} \label{eq:mgder1-3} \\
&\dfrac{\langle \epsilon, \der \TT{+k =d pred}, \MT{\lambda P. \lambda Q. Q(P)} \rangle
    \langle \TT{eat}, \der \TT{-f}, \MT{\lambda x. \lambda y. eat(x)(y)} \rangle
    \langle \TT{cheese}, \der \TT{-k}, \MT{cheese} \rangle
    }
    {\langle \TT{cheese}, \der \TT{=d pred}, \MT{(\lambda P. \lambda Q. Q(P))(cheese)} \rangle
    \langle \TT{eat}, \der \TT{-f}, \MT{\lambda x. \lambda y. eat(x)(y)} \rangle
    } \: \text{move-1} \label{eq:mgder1-4} \\
&\dfrac{\langle \TT{cheese}, \der \TT{=d pred}, \MT{(\lambda P. \lambda Q. Q(P))(cheese)} \rangle
    \langle \TT{eat}, \der \TT{-f}, \MT{\lambda x. \lambda y. eat(x)(y)} \rangle
    }
    {\langle \TT{cheese}, \der \TT{=d pred}, \MT{\lambda Q. Q(cheese)} \rangle
    \langle \TT{eat}, \der \TT{-f}, \MT{\lambda x. \lambda y. eat(x)(y)} \rangle
    } \: \lambda\text{-app.} \label{eq:mgder1-5} \\
&\dfrac{\langle \TT{cheese}, \der \TT{=d pred}, \MT{\lambda Q. Q(cheese)} \rangle
    \langle \TT{eat}, \der \TT{-f}, \MT{\lambda x. \lambda y. eat(x)(y)} \rangle
    \qquad
    \langle \TT{the mouse}, \der \TT{d -k}, \MT{mouse}\rangle
    }
    {\langle \TT{cheese}, \der \TT{pred}, \MT{\lambda Q. Q(cheese)} \rangle
    \langle \TT{eat}, \der \TT{-f}, \MT{\lambda x. \lambda y. eat(x)(y)} \rangle
    \langle \TT{the mouse}, \der \TT{-k}, \MT{mouse}\rangle
    } \: \text{merge-3} \label{eq:mgder1-6} \\
&\dfrac{\langle \TT{-s}, \lex \TT{=pred +f +k t}, \epsilon \rangle
    \qquad
    \langle \TT{cheese}, \der \TT{pred}, \MT{\lambda Q. Q(cheese)} \rangle
    \langle \TT{eat}, \der \TT{-f}, \MT{\lambda x. \lambda y. eat(x)(y)} \rangle
    \langle \TT{the mouse}, \der \TT{-k}, \MT{mouse}\rangle
    }
    {\langle \TT{-s cheese}, \der \TT{+f +k t}, \MT{\lambda Q. Q(cheese)}\rangle
    \langle \TT{eat}, \der \TT{-f}, \MT{\lambda x. \lambda y. eat(x)(y)} \rangle
    \langle \TT{the mouse}, \der \TT{-k}, \MT{mouse}\rangle
    } \: \text{merge-1} \label{eq:mgder1-7} \\
&\dfrac{\langle \TT{-s cheese}, \der \TT{+f +k t}, \MT{\lambda Q. Q(cheese)}\rangle
    \langle \TT{eat}, \der \TT{-f}, \MT{\lambda x. \lambda y. eat(x)(y)} \rangle
    \langle \TT{the mouse}, \der \TT{-k}, \MT{mouse}\rangle
    }
    {
    \langle \TT{eat-s cheese}, \der \TT{+k t}, \MT{(\lambda Q. Q(cheese))(\lambda x. \lambda y. eat(x)(y))}\rangle
    \langle \TT{the mouse}, \der \TT{-k}, \MT{mouse}\rangle
    } \: \text{move-1} \label{eq:mgder1-8} \\
&\dfrac{\langle \TT{eats cheese}, \der \TT{+k t}, \MT{(\lambda Q. Q(cheese))(\lambda x. \lambda y. eat(x)(y))}\rangle
    \langle \TT{the mouse}, \der \TT{-k}, \MT{mouse}\rangle
    }
    {\langle \TT{eats cheese}, \der \TT{+k t}, \MT{(\lambda x. \lambda y. eat(x)(y))(cheese)}\rangle
    \langle \TT{the mouse}, \der \TT{-k}, \MT{mouse}\rangle
    } \: \lambda\text{-app.} \label{eq:mgder1-9} \\
&\dfrac{\langle \TT{eats cheese}, \der \TT{+k t}, \MT{(\lambda x. \lambda y. eat(x)(y))(cheese)}\rangle
    \langle \TT{the mouse}, \der \TT{-k}, \MT{mouse}\rangle
    }
    {\langle \TT{eats cheese}, \der \TT{+k t}, \MT{\lambda y. eat(cheese)(y)}\rangle
    \langle \TT{the mouse}, \der \TT{-k}, \MT{mouse}\rangle
    } \: \lambda\text{-app.} \label{eq:mgder1-10} \\
&\dfrac{\langle \TT{eats cheese}, \der \TT{+k t}, \MT{\lambda y. eat(cheese)(y)}\rangle
    \langle \TT{the mouse}, \der \TT{-k}, \MT{mouse}\rangle
    }
    {
    \langle \TT{the mouse eats cheese}, \der \TT{t}, \MT{(\lambda y. eat(cheese)(y))(mouse)}\rangle
    } \: \text{move-1} \label{eq:mgder1-11} \\
&\dfrac{\langle \TT{the mouse eats cheese}, \der \TT{ t}, \MT{(\lambda y. eat(cheese)(y))(mouse)}\rangle
    }
    {\langle \TT{the mouse eats cheese}, \der \TT{t}, \MT{eat(cheese)(mouse)}\rangle
    } \: \lambda\text{-app.} \label{eq:mgder1-12} \\
&\dfrac{\langle \epsilon, \lex \TT{=t c}, \epsilon \rangle
    \qquad
    \langle \TT{the mouse eats cheese}, \der \TT{t}, \MT{eat(cheese)(mouse)}\rangle
    }
    {\langle \TT{the mouse eats cheese}, \der \TT{c}, \MT{eat(cheese)(mouse)}\rangle
    } \: \text{merge-1} \label{eq:mgder1-13} \:.
\end{align}
\end{subequations}
}

In the first step, \eqref{eq:mgder1-1}, the determiner \TT{the} takes the noun \TT{mouse} as its complement to form a determiner phrase \TT{d} that requires licensing through case marking afterwards. In step 2, the finite verb \TT{eat} selects the noun \TT{cheese} as direct object, thus forming a verbal phrase \TT{v}. As there remain unchecked features, only merge-3 applies yielding a minimalist expression, i.e. a sequence of signs. In step 3, the phonetically empty predicate \TT{pred} merges with the formerly built verbal phrase. Since \TT{pred} assigns accusative case, the direct object is moved in \eqref{eq:mgder1-4} toward the first position through case marking by simultaneously concatenating the respective lambda terms. Thus, lambda application \index{lambda application} entails the expression in step 5. Then, in step 6, the predicate selects its subject, the formerly construed determiner phrase. In the seventh step, \eqref{eq:mgder1-7}, the present-tense suffix unifies with the predicate, entailing an inflection phrase \TT{pred}, whose verb is moved into the first position in step 8, thereby yielding the inflected verb \TT{eat-s}. In steps 9 and 10 two lambda applications result into the correct semantics, already displayed in \Fig{fig:gb}(b). Step 11 assigns nominative case to the subject through movement into specifier position. A further lambda application \index{lambda application} in step 12 yields the intended interpretation of predicate logics. Finally, in step 13, the syntactic type \TT{t} is converted into \TT{c} to obtain the proper start symbol of the grammar\index{grammar}.


\subsection{Utterance-Meaning Transducer} \index{utterance-meaning transducer (UMT)}
\label{sec:umt}

Derivations such as \eqref{eq:mgder1} are essential for minimalist grammar. \index{grammar| minimalist grammar} However, their computation is neither incremental nor predictive. Therefore, they are not suitable for natural language processing in their present form of data-driven bottom-up processing. A number of different parsing \index{parsing} architectures have been suggested in the literature to remedy this problem \cite{Harkema01, Mainguy10, Stabler11b, StanojevicStabler18}. From a psycholinguistic point of view, predictive parsing \index{parsing} appears most plausible, because a cognitive agent should be able to make informed guesses about a speaker's intents as early as possible, without waiting for the end of an utterance \cite{Hale11}. This makes either an hypothesis-driven top-down parser, or a mixed-strategy left-corner parser desirable also for language engineering applications. \index{lambda application} In this section, we briefly describe a bidirectional utterance-meaning transducer (UMT) for MG that is based upon Stabler's top-down recognizer \cite{Stabler11b} as outlined earlier in \cite{GrabenMeyerEA19}. Its generalization towards the recent left-corner parser \cite{StanojevicStabler18} is straightforward.

\begin{figure}[H]
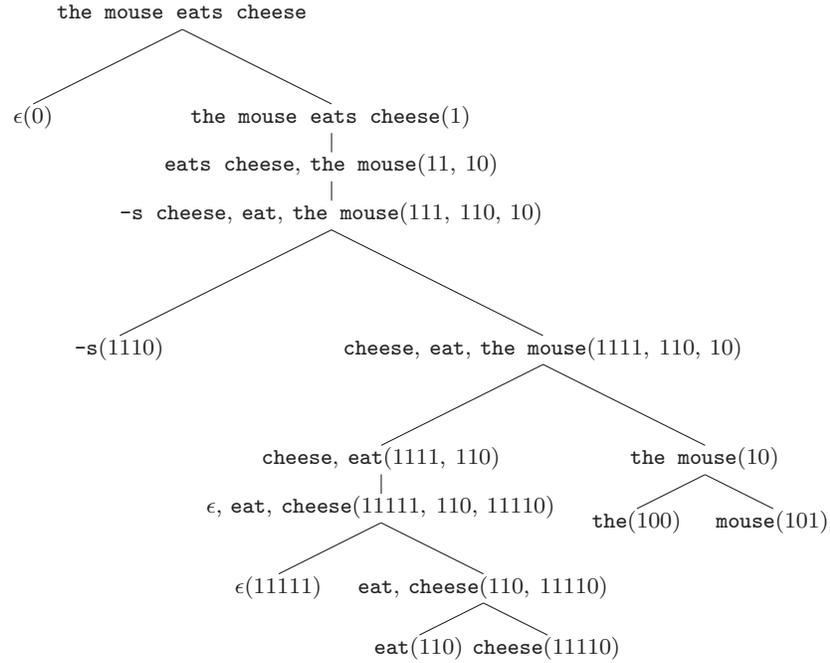
\small
\Tree [.\TT{the mouse eats cheese}
            $\epsilon$(0) !\qsetw{1cm}
            [.\TT{the mouse eats cheese}(1)
                [.{\TT{eats cheese}, \TT{the mouse}(11, 10)}
                    [.{\TT{-s cheese}, \TT{eat}, \TT{the mouse}(111, 110, 10)}
                        \TT{-s}(1110) !\qsetw{1cm}
                        [.{\TT{cheese}, \TT{eat}, \TT{the mouse}(1111, 110, 10)}
                            [.{\TT{cheese}, \TT{eat}(1111, 110)} !\qsetw{2.2cm}
                                [.{$\epsilon$, \TT{eat}, \TT{cheese}(11111, 110, 11110)}
                                    $\epsilon$(11111) !\qsetw{2cm}
                                    [.{\TT{eat}, \TT{cheese}(110, 11110)}
                                        \TT{eat}(110) \TT{cheese}(11110)
                                    ].{\TT{eat}, \TT{cheese}(110, 11110)}
                                ].{$\epsilon$, \TT{eat}, \TT{cheese}(11111, 110, 11110)}
                            ].{\TT{cheese}, \TT{eat}(1111, 110)} !\qsetw{5.5cm}
                            [.\TT{the mouse}(10)
                                \TT{the}(100) !\qsetw{2cm}
                                \TT{mouse}(101)
                            ].\TT{the mouse}(10)
                        ].{\TT{cheese}, \TT{eat}, \TT{the mouse}(1111, 110, 10)}
                    ].{\TT{-s cheese}, \TT{eat}, \TT{the mouse}(111, 110, 10)}
                ].{\TT{eats cheese}, \TT{the mouse}(11, 10)}
            ].\TT{the mouse eats cheese}(1)
    ].\TT{the mouse eats cheese}
\caption{\label{fig:der1tree} Simplified derivation tree of \eqref{eq:mgder1}. Exponents of different signs are separated by commas. Nodes are also addressed by index tuples.}
\end{figure}

The central object for MG language processing is the \emph{derivation tree} obtained from a bottom-up derivation as in \eqref{eq:mgder1}. Figure \ref{fig:der1tree} depicts this derivation tree, where we present a comma-separated sequence of exponents \index{exponent} for the sake of simplicity. Additionally, every node is addressed by an index tuple that is computed according to Stabler's algorithm \cite{Stabler11b}.

Pursuing the tree paths in \Fig{fig:der1tree} from the bottom to the top, provides exactly the derivation \eqref{eq:mgder1}. However, reading it from the top towards the bottom allows for an interpretation in terms of \emph{multiple context-free grammars} \cite{SekiEA91, Michaelis01a}\index{grammar | multiple context-free} (MCFG) where categories are $n$-ary predicates over string exponents. \index{exponent} Like in context-free grammars, \index{grammar | context-free} every branching in the derivation tree \Fig{fig:der1tree} leads to one phrase structure rule in the MCFG. Thus, the MCFG enumerated in \eqref{eq:mcfg1} codifies the MG \Tab{tab:mg0}.

{\scriptsize
\begin{subequations}\label{eq:mcfg1}
\renewcommand{\theequation}{\theparentequation{}-{}\arabic{equation}}
\begin{align}
    &\langle \der \TT{c}\rangle(e_0 e_1) \gets \langle \lex \TT{=t c} \rangle(e_0) \quad \langle \der \TT{t} \rangle(e_1) \label{eq:mcfg1-1} \\
    &\langle \der \TT{t}\rangle(e_1 e_0) \gets \langle \der \TT{+k t, -k}\rangle(e_0, e_1) \label{eq:mcfg1-2} \\
    &\langle \der \TT{+k t, -k}\rangle(e_1 e_0, e_2) \gets \langle \der \TT{+f +k t, -f, -k}\rangle(e_0, e_1, e_2) \label{eq:mcfg1-3} \\
    &\langle \der \TT{+f +k t, -f, -k}\rangle(e_0 e_1, e_2, e_3) \gets
    \langle \lex \TT{=pred +f +k t} \rangle(e_0) \quad \langle \der \TT{pred, -f, -k}\rangle(e_1, e_2, e_3) \label{eq:mcfg1-4} \\
    &\langle \der \TT{pred, -f, -k}\rangle(e_0, e_1, e_2) \gets \langle \der \TT{=d pred, -f} \rangle(e_0, e_1) \quad
    \langle \der \TT{d -k} \rangle(e_2) \label{eq:mcfg1-5} \\
    &\langle \der \TT{=d pred, -f}\rangle(e_2 e_0, e_1) \gets \langle \der \TT{+k =d pred, -f, -k}\rangle(e_0, e_1, e_2) \label{eq:mcfg1-6} \\
    &\langle \der \TT{+k =d pred, -f, -k}\rangle(e_0, e_1, e_2) \gets \langle \lex \TT{=v +k =d pred}\rangle(e_0)  \quad
    \langle \der \TT{v -f, -k}\rangle(e_1, e_2) \label{eq:mcfg1-7} \\
    &\langle \der \TT{v -f, -k}\rangle(e_0, e_1) \gets \langle \lex \TT{=n v -f}\rangle(e_0) \quad \langle \lex \TT{n -k}\rangle(e_1) \label{eq:mcfg1-8} \\
    &\langle \der \TT{d -k}\rangle(e_0 e_1) \gets \langle \lex \TT{=n d -k} \rangle(e_0) \quad \langle \lex \TT{n} \rangle(e_1)     \label{eq:mcfg1-9} \\
    &\langle \lex \TT{n}\rangle(\TT{mouse}) \label{eq:mcfg1-10} \\
    &\langle \lex \TT{n -k} \rangle(\TT{cheese}) \label{eq:mcfg1-11} \\
    &\langle \lex \TT{=n d -k} \rangle(\TT{the}) \label{eq:mcfg1-12} \\
    &\langle \lex \TT{=n v -f}\rangle(\TT{eat}) \label{eq:mcfg1-13} \\
    &\langle \lex \TT{=pred +f +k t}\rangle(\TT{-s}) \label{eq:mcfg1-14} \\
    &\langle \lex \TT{=v +k =d pred}\rangle(\epsilon) \label{eq:mcfg1-15} \\
    &\langle \lex \TT{=t c}\rangle(\epsilon) \label{eq:mcfg1-16}
\end{align}
\end{subequations}
}

In \eqref{eq:mcfg1}, the angular brackets enclose the MCFG categories that are obviously formed by tuples of MG categories and syntactic types. These categories have the same number of string arguments $e_k$ as prescribed in the type tuples. Because MCFG serve only for syntactic parsing \index{parsing} in our UMT, \index{utterance-meaning transducer (UMT)} we deliberately omit the semantic terms here; they are reintroduced below. The MCFG rules (\ref{eq:mcfg1-1} --  \ref{eq:mcfg1-9}) are directly obtained from the derivation tree \Fig{fig:der1tree} by reverting the merge and move operations of \eqref{eq:mgder1} through their ``unmerge'' and ``unmove'' counterparts \cite{Harkema01}. The MCFG axioms, i.e. the lexical rules (\ref{eq:mcfg1-10} -- \ref{eq:mcfg1-16}), are reformulations of the entries in the MG lexicon \Tab{tab:mg0}.


\subsubsection{Language Production}
\label{sec:prod}

The UMT's \index{utterance-meaning transducer (UMT)} language production module finds a semantic representation of an intended utterance in form of a Sch\"onfinkel-Curry \cite{Lohnstein11, Schonfinkel24} formula of predicate logic, such as $\MT{eat(cheese)(mouse)}$, for instance. According to \Fig{fig:gb}(b) this is a hierarchical data structure that can control the MG derivation \eqref{eq:mgder1}. Thus, the cognitive agent accesses its mental lexicon, either through \Tab{tab:mg0} or its MCFG twin \eqref{eq:mcfg1} in order to retrieve the linguistic signs \index{linguistic sign} for the denotations \TT{eat}, \TT{cheese}, and \TT{mouse}. Then, the semantic tree \Fig{fig:gb}(b) governs the correct derivation \eqref{eq:mgder1} up to lexicon entries that are phonetically empty. These must occasionally be queried from the data base whenever required. At the end of the derivation the computed exponent \index{exponent} \TT{the mouse eats cheese} is uttered.


\subsubsection{Language Understanding}
\label{sec:under}

The language understanding module of our UMT \index{utterance-meaning transducer (UMT)} comprises three memory tapes: the input sequence, a syntactic priority queue, and also a semantic priority queue. Input tape and syntactic priority queue together constitute Stabler's priority queue top-down parser \cite{Stabler11b}. Yet, in order to compute the meaning of an utterance in the semantic priority queue, we slightly depart from the original proposal by omitting the simplifying trim function. Table \ref{tab:tdr} presents the temporal evolution of the top-down recognizer's configurations while processing the utterance \TT{the mouse eats cheese}.

\begin{table}[H]
\centering \scriptsize
\caption{\label{tab:tdr} MG top-down parse of \TT{the mouse eats cheese}.}
\begin{tabular}{rlll}
  \hline
  step & input & syntactic queue & operation \\
  \hline
  1. & \TT{the mouse eats cheese} & {$\langle \der \TT{c}\rangle(\epsilon)$} & expand \eqref{eq:mcfg1-1} \\
  2. & \TT{the mouse eats cheese} & {$\langle \lex \TT{=t c} \rangle(0)\langle \der \TT{t} \rangle(1) $} & scan \eqref{eq:mcfg1-16} \\
  3. & \TT{the mouse eats cheese} & {$\langle \der \TT{t} \rangle(1)$} & expand \eqref{eq:mcfg1-2} \\
  4. & \TT{the mouse eats cheese} & {$\langle \der \TT{+k t, -k}\rangle(11, 10)$} & expand \eqref{eq:mcfg1-3} \\
  5. & \TT{the mouse eats cheese} & {$\langle \der \TT{+f +k t, -f, -k}\rangle(111, 110, 10)$} & expand \eqref{eq:mcfg1-4} \\
  6. & \TT{the mouse eats cheese} & {$\langle \lex \TT{=pred +f +k t} \rangle(1110) \langle \der \TT{pred, -f, -k}\rangle(1111, 110, 10)$} & sort \\
  7. & \TT{the mouse eats cheese} & {$\langle \der \TT{pred, -f, -k}\rangle(1111, 110, 10)\langle \lex \TT{=pred +f +k t} \rangle(1110) $} & expand \eqref{eq:mcfg1-5} \\
  8. & \TT{the mouse eats cheese} & {$\langle \der \TT{=d pred, -f} \rangle(1111, 110)
    \langle \der \TT{d -k} \rangle(10) \langle \lex \TT{=pred +f +k t} \rangle(1110) $} & sort \\
  9. & \TT{the mouse eats cheese} & {$\langle \der \TT{d -k} \rangle(10) \langle \der \TT{=d pred, -f} \rangle(1111, 110)
    \langle \lex \TT{=pred +f +k t} \rangle(1110) $} & expand \eqref{eq:mcfg1-9}  \\
  10. & \TT{the mouse eats cheese} & {$\langle \lex \TT{=n d -k} \rangle(100) \langle \lex \TT{n} \rangle(101)
  \langle \der \TT{=d pred, -f} \rangle(1111, 110) \langle \lex \TT{=pred +f +k t} \rangle(1110) $} & scan \eqref{eq:mcfg1-12} \\
  11. & \TT{mouse eats cheese} & {$\langle \lex \TT{n} \rangle(101)
  \langle \der \TT{=d pred, -f} \rangle(1111, 110) \langle \lex \TT{=pred +f +k t} \rangle(1110) $} & scan \eqref{eq:mcfg1-10} \\
  12. & \TT{eats cheese} & {$\langle \der \TT{=d pred, -f} \rangle(1111, 110) \langle \lex \TT{=pred +f +k t} \rangle(1110) $} & expand \eqref{eq:mcfg1-6} \\
  13. & \TT{eats cheese} & {$\langle \der \TT{+k =d pred, -f, -k}\rangle(11111, 110, 11110) \langle \lex \TT{=pred +f +k t} \rangle(1110) $} & expand \eqref{eq:mcfg1-7} \\
  14. & \TT{eats cheese} & {$\langle \lex \TT{=v +k =d pred}\rangle(11111)
  \langle \der \TT{v -f, -k}\rangle(110, 11110) \langle \lex \TT{=pred +f +k t} \rangle(1110) $} & sort \\
  15. & \TT{eats cheese} & {$\langle \der \TT{v -f, -k}\rangle(110, 11110) \langle \lex \TT{=pred +f +k t} \rangle(1110) \langle \lex \TT{=v +k =d pred}\rangle(11111)$} &  expand \eqref{eq:mcfg1-8} \\
  16. & \TT{eats cheese} & {$\langle \lex \TT{=n v -f}\rangle(110) \langle \lex \TT{n -k}\rangle(11110)\langle \lex \TT{=pred +f +k t} \rangle(1110) \langle \lex \TT{=v +k =d pred}\rangle(11111)$} &  sort \\
  17. & \TT{eats cheese} & {$\langle \lex \TT{=n v -f}\rangle(110) \langle \lex \TT{=pred +f +k t} \rangle(1110) \langle \lex \TT{n -k}\rangle(11110)\langle \lex \TT{=v +k =d pred}\rangle(11111)$} &  scan \eqref{eq:mcfg1-13} \\
  18. & \TT{-s cheese} & {$\langle \lex \TT{=pred +f +k t} \rangle(1110) \langle \lex \TT{n -k}\rangle(11110)\langle \lex \TT{=v +k =d pred}\rangle(11111)$} &  scan \eqref{eq:mcfg1-14} \\
  19. & \TT{cheese} & {$\langle \lex \TT{n -k}\rangle(11110)\langle \lex \TT{=v +k =d pred}\rangle(11111)$} &  scan \eqref{eq:mcfg1-11} \\
  20. & $\epsilon$ & {$\langle \lex \TT{=v +k =d pred}\rangle(11111)$} &  scan \eqref{eq:mcfg1-15} \\
  21. & $\epsilon$ & $\epsilon$ & accept \\
  \hline
\end{tabular}
\end{table}

The parser is initialized with the input string to be processed and the MCFG start symbol $\langle \der \TT{c}\rangle(\epsilon)$ --- corresponding to the MG start symbol \TT{c} --- at the top of the priority queue. For each rule of the MCFG \eqref{eq:mcfg1}, the algorithm replaces its string variables by an index tuple that addresses the corresponding nodes in the derivation tree \Fig{fig:der1tree} \cite{Stabler11b}. These indices allow for an ordering relation where shorter indices are smaller than longer ones, while indices of equal length are ordered lexicographically. As a consequence, the MCFG axioms in \eqref{eq:mcfg1} become ordered according to their temporal appearance in the utterance. Using the notation of the derivation tree \Fig{fig:der1tree}, we get
\[
    \TT{the}(100) < \TT{mouse}(101) < \TT{eat}(110) < \TT{-s}(1110) < \TT{cheese}(11110) \:.
\]
Hence, index sorting ensures incremental parsing. \index{parsing}

Besides the occasional sorting of the syntactic priority queue, the automaton behaves as a conventional context-free top-down parser. When the first item in the queue is an MCFG category appearing at the left hand side of an MCFG rule, this item is \emph{expand}ed into the right hand side of that rule. When the first item in the queue is a predicted MCFG axiom whose exponent \index{exponent} also appears on top of the input tape, this item is \emph{scan}ned from the input and thereby removed from queue and input simultaneously. Finally, if queue and input both contain only the empty word, the utterance has been successfully recognized and the parser terminates in the \emph{accept}ing state.

Interestingly, the index formalism leads to a straightforward implementation of the UMT's \index{utterance-meaning transducer (UMT)} semantic parser \index{semantic parser} as well. The derivation tree \Fig{fig:der1tree} reveals that the index length correlates with the tree depth. In our example, the items
$\langle \lex \TT{n -k}\rangle(11110)$ and $\langle \lex \TT{=v +k =d pred}\rangle(11111)$ in the priority queue have the longest indices. These correspond precisely to the lambda terms $\MT{cheese}$ and $\MT{\lambda P. \lambda Q. Q(P)}$, respectively, that are unified by lambda application \index{lambda application} in derivation step \eqref{eq:mgder1-5}. Moreover, also the semantic analysis in \Fig{fig:gb}(b) illustrates that the deepest nodes are semantically unified first.

Every time, when the syntactic parser scans a correctly predicted item from the input tape, this item is removed from both input tape and syntactic priority queue. Simultaneously, the semantic content of its sign is pushed on top of the semantic priority queue, yet preserving its index. When some or all semantic items are stored in the queue, they are sorted in \emph{reversed} index order to get highest semantic priority on top of the queue. Table \ref{tab:semp} illustrates the semantic parsing \index{parsing} for the given example.

 \begin{table}[H]
\centering \scriptsize
\caption{\label{tab:semp} Semantic processing of \TT{the mouse eats cheese}.}
\begin{tabular}{rlll}
  \hline
  step & input & semantic queue & operation \\
  \hline
  1. & \TT{the mouse eats cheese} & {$\epsilon$} & scan \eqref{eq:mcfg1-16} \\
  2. & \TT{the mouse eats cheese} & {$\langle \epsilon\rangle(0)$} & apply \\
  3. & \TT{the mouse eats cheese} & {$\epsilon$} & scan \eqref{eq:mcfg1-12} \\
  4. & \TT{mouse eats cheese} & {$\langle \epsilon \rangle(100)$} & apply\\
  5. & \TT{mouse eats cheese} & {$\epsilon$} & scan \eqref{eq:mcfg1-10} \\
  6. & \TT{eats cheese} & {$\langle \MT{mouse}\rangle(101)$} & scan \eqref{eq:mcfg1-13} \\
  7. & \TT{-s cheese} & {$\langle \MT{mouse}\rangle(101)\langle \MT{\lambda x. \lambda y. eat(x)(y)}\rangle(110)$} &  scan \eqref{eq:mcfg1-14} \\
  8. & \TT{cheese} & {$\langle \MT{mouse}\rangle(101)\langle \MT{\lambda x. \lambda y. eat(x)(y)}\rangle(110)\langle \epsilon \rangle(1110) $} &  apply \\
  9. & \TT{cheese} & {$\langle \MT{mouse}\rangle(101)\langle \MT{\lambda x. \lambda y. eat(x)(y)}\rangle(110)$} &  scan \eqref{eq:mcfg1-11} \\
  10. & $\epsilon$ & {$\langle \MT{mouse}\rangle(101)\langle \MT{\lambda x. \lambda y. eat(x)(y)}\rangle(110)\langle \MT{cheese}\rangle(11110)$} &  scan \eqref{eq:mcfg1-15} \\
  11. & $\epsilon$ & {$\langle \MT{mouse}\rangle(101)\langle \MT{\lambda x. \lambda y. eat(x)(y)}\rangle(110)\langle \MT{cheese}\rangle(11110)\langle \MT{\lambda P. \lambda Q. Q(P)}\rangle(11111)$} &  sort \\
  11. & $\epsilon$ & {$\langle \MT{\lambda P. \lambda Q. Q(P)}\rangle(11111) \langle \MT{cheese}\rangle(11110)
  \langle \MT{\lambda x. \lambda y. eat(x)(y)}\rangle(110) \langle \MT{mouse}\rangle(101)$} &  apply \\
  12. & $\epsilon$ & {$\langle \MT{\lambda Q. Q(cheese)}\rangle(1111)
  \langle \MT{\lambda x. \lambda y. eat(x)(y)}\rangle(110) \langle \MT{mouse}\rangle(101)$} &  apply \\
  13. & $\epsilon$ & {$\langle \MT{(\lambda x. \lambda y. eat(x)(y))(cheese)}\rangle(111) \langle \MT{mouse}\rangle(101)$} &  apply \\
  14. & $\epsilon$ & {$\langle \MT{\lambda y. eat(cheese)(y)}\rangle(111) \langle \MT{mouse}\rangle(101)$} &  apply \\
  15. & $\epsilon$ & {$\langle \MT{eat(cheese)(mouse)}\rangle(11) $} &  understand \\
  \hline
\end{tabular}
\end{table}

In analogy to the syntactic recognizer, the semantic parser \index{semantic parser} operates in similar modes. Both processors share their common \emph{scan} operation. In contrast to the syntactic parser which sorts indices in ascending order, the semantic module \emph{sort}s them in descending order for operating on the deepest nodes in the derivation tree first. Most of the time, it attempts lambda application \index{lambda application}  (\emph{apply}) which is always preferred for $\epsilon$-items on the queue. When apply has been sufficiently performed to a term, the last index number is removed from its index (sometimes it might also be necessary to exchange two items for lambda application). Finally, the semantic parser \index{semantic parser} terminates in the \emph{understand}ing state.


\section{Reinforcement Learning}
\label{sec:ril}

Sofar we discussed how a cognitive agent, being either human or an intelligent machine, could produce and understand utterances that are described in terms of minimalist grammar. \index{grammar| minimalist grammar} An MG is given by a mental lexicon as in example \Tab{tab:mg0}, encoding a large amount of linguistic expert knowledge. Therefore, it seems unlikely that speech-controlled user interfaces \index{user interface | speech-controlled user interface} could be build and sold by engineering companies for little expenses.

Yet, it has been shown that MG are effectively learnable in the sense of Gold's formal learning theory \cite{Gold67}\index{learning theory}. The studies \cite{BonatoRetore01, KobeleCollierEA02, StablerEA03} demonstrated how MG can be acquired by positive examples from linguistic dependence graphs \cite{BostonHaleKuhlmann10, Nivre03}. The required dependency structures can be extracted from linguistic corpora by means of big data machine learning \index{machine learning} techniques, such as the expectation maximization (EM) algorithm \cite{KleinManning04}.

In our terminology, such statistical learning methods only consider correlations at the exponent \index{exponent} level of linguistic signs. \index{linguistic sign} By contrast, in the present study we propose an alternative training algorithm that simultaneously analyzes similarities between exponents \index{exponent} and semantic terms. Moreover, we exploit both positive and negative examples to obtain a better performance through reinforcement learning \cite{Skinner15, SuttonBarto18}\index{reinforcement learning}.

The language learner is a cognitive agent $\mathcal{L}$ in a state $X_t$, to be identified with $\mathcal{L}$'s mental lexicon at training time $t$. At time $t = 0$, $\mathcal{L}$ is initialized as a \emph{tabula rasa} with empty lexicon
\begin{equation}\label{eq:lexini}
    X_0 \gets \emptyset
\end{equation}
and exposed to UMPs \index{utterance meaning pair (UMP)} produced by a teacher $\mathcal{T}$. Note that we assume $\mathcal{T}$ presenting already complete UMPs \index{utterance meaning pair (UMP)} and not singular utterances to $\mathcal{L}$. Thus we circumvent the \emph{symbol grounding problem} of firstly assigning meanings $\sigma$ to uttered exponents \index{exponent} $e$ \cite{Harnad90}, which will be addressed in future research. Moreover, we assume that $\mathcal{L}$ is instructed to reproduce $\mathcal{T}$'s utterances based on its own semantic understanding. This provides a feedback loop and therefore applicability of reinforcement learning \cite{Skinner15, SuttonBarto18}\index{reinforcement learning}. For our introductory example, we adopt the simple semantic model from \Sec{sec:mg}. In each iteration, the teacher utters an UMP \index{utterance meaning pair (UMP)} that should be learned by the learner.


\paragraph{\em First iteration}
\label{sec:it1}

Let the teacher $\mathcal{T}$ make the first utterance \eqref{eq:ump1}
\[
    u_1 = \langle \TT{the mouse eats cheese}, \MT{eat(cheese)(mouse)} \rangle \:.
\]

As long as $\mathcal{L}$ is not able to detect patterns or common similarities in $\mathcal{T}$'s UMPs, \index{utterance meaning pair (UMP)} it simply adds new entries directly to its mental lexicon, assuming that all utterances are complex ``\der'' and possessing base type \TT{c}, i.e. the MG start symbol. Hence, $\mathcal{L}$'s state $X_t$ evolves according to the update rule
\begin{equation}\label{eq:dyn1}
    X_t \gets X_{t - 1} \cup \{ \langle e_t, \der \TT{c}, \sigma_t \rangle \} \:,
\end{equation}
when $u_t = \langle e_t, \sigma_t \rangle$ is the UMP \index{utterance meaning pair (UMP)} presented at time $t$ by $\mathcal{T}$.

In this way, the mental lexicon $X_1$ shown in \Tab{tab:mg1} has been acquired at time $t = 1$.

\begin{table}[H]
\caption{\label{tab:mg1} Learned minimalist lexicon $X_1$ at time $t = 1$.}
\[
\begin{array}{ll}
    \langle \TT{the mouse eats cheese}, \der \TT{c}, \MT{eat(cheese)(mouse)} \rangle
\end{array}
\]
\end{table}


\paragraph{\em Second iteration}
\label{sec:it2}

Next, let the teacher utter another proposition
\begin{equation}\label{eq:ump2}
    u_2 = \langle \TT{the rat eats cheese}, \MT{eat(cheese)(rat)} \rangle \:.
\end{equation}
Looking at $u_1, u_2$ together, the agent's pattern matching module is able to find similarities between exponents \index{exponent} and semantics, underlined in \eqref{eq:patmat1}.
\begin{subequations}\label{eq:patmat1}
\renewcommand{\theequation}{\theparentequation{}-{}\arabic{equation}}
\begin{align}
    u_1 &= \langle \TT{\underline{the} mouse \underline{eats cheese}}, \MT{\underline{eat(cheese)}(mouse)} \rangle \\
    u_2 &= \langle \TT{\underline{the} rat \underline{eats cheese}}, \MT{\underline{eat(cheese)}(rat)} \rangle \:.
\end{align}
\end{subequations}

Thus, $\mathcal{L}$ creates two distinct items for \TT{the mouse} and \TT{the rat}, respectively, and carries out lambda abstraction \index{lambda abstraction} to obtain the updated lexicon $X_2$ in \Tab{tab:mg2.1}.

\begin{table}[H]
\caption{\label{tab:mg2.1} Learned minimalist lexicon $X_2$ at time $t = 2$.}
\[
\begin{array}{ll}
    \langle \TT{\underline{the} mouse}, \der \TT{d}, \MT{mouse} \rangle &
    \langle \TT{\underline{the} rat}, \der \TT{d}, \MT{rat} \rangle \\
    \langle \TT{eats cheese}, \der \TT{=d c}, \MT{\lambda y.eat(cheese)(y)} \rangle
\end{array}
\]
\end{table}
Note that the induced variable symbol \TT{y} and syntactic types \TT{d}, \TT{c} are completely arbitrary and do not have any particular meaning to the agent.

As indicated by underlines in \Tab{tab:mg2.1}, the exponents \index{exponent} \TT{the mouse} and \TT{the rat}, could be further segmented through pattern matching, that is not reflected by their semantic counterparts, though. Therefore, a revised lexicon $X_{21}$, displayed in \Tab{tab:mg2.2} can be devised.

\begin{table}[H]
\caption{\label{tab:mg2.2} Revised minimalist lexicon $X_{21}$.}
\[
\begin{array}{ll}
    \langle \TT{the}, \lex \TT{=n d}, \epsilon \rangle &
    \langle \TT{mouse}, \lex \TT{n}, \MT{mouse} \rangle \\
    \langle \TT{rat}, \lex \TT{n}, \MT{rat} \rangle &
    \langle \TT{eats cheese}, \der \TT{=d c}, \MT{\lambda y.eat(cheese)(y)} \rangle
\end{array}
\]
\end{table}

For closing the reinforcement cycle, $\mathcal{L}$ is supposed to produce utterances upon its own understanding. Thus, we assume that $\mathcal{L}$ wants to express the proposition $\MT{eat(cheese)(rat)}$. According to our discussion in \Sec{sec:prod}, the corresponding signs are retrieved from the lexicon $X_{21}$ and processed through a valid derivation leading to the correct utterance \TT{the rat eats cheese}, that is subsequently endorsed by $\mathcal{T}$.


\paragraph{\em Third iteration}
\label{sec:it3}

In the third training session, the teacher's utterance might be
\begin{equation}\label{eq:ump3}
    u_3 = \langle \TT{the mouse eats carrot}, \MT{eat(carrot)(mouse)} \rangle \:.
\end{equation}

Now we have to compare $u_3$ with the lexicon entry for \TT{eats cheese} in \eqref{eq:patmat2}.
\begin{subequations}\label{eq:patmat2}
\renewcommand{\theequation}{\theparentequation{}-{}\arabic{equation}}
\begin{align}
    &\langle \TT{the mouse \underline{eats} carrot}, \MT{\underline{eat}(carrot)(mouse)} \rangle \\
    & \langle \TT{\underline{eats} cheese}, \der \TT{c}, \MT{\lambda y.\underline{eat}(cheese)(y)} \rangle \:.
\end{align}
\end{subequations}

Another lambda abstraction \index{lambda abstraction} entails the lexicon $X_3$ in \Tab{tab:mg3}.

\begin{table}[H]
\caption{\label{tab:mg3} Learned minimalist lexicon $X_3$ at time $t = 3$.}
\[
\begin{array}{ll}
    \langle \TT{the}, \lex \TT{=n d}, \epsilon \rangle &
    \langle \TT{mouse}, \lex \TT{n}, \MT{mouse} \rangle \\
    \langle \TT{rat}, \lex \TT{n}, \MT{rat} \rangle &
    \langle \TT{cheese}, \lex \TT{n}, \MT{cheese} \rangle \\
    \langle \TT{carrot}, \lex \TT{n}, \MT{carrot} \rangle &
    \langle \TT{eats}, \lex \TT{=n =d c}, \MT{\lambda x.\lambda y.eat(x)(y)} \rangle
\end{array}
\]
\end{table}

Here, the learner assumes that \TT{eats} is a simple, lexical category without having further evidence as in \Sec{sec:mg}.

Since $\mathcal{L}$ is instructed to produce well-formed utterances, it could now generate a novel semantic representation, such as $\MT{eat(carrot)(rat)}$. This leads through data base query from the mental lexicon $X_3$ to the correct derivation \eqref{eq:mgder2} that is rewarded by $\mathcal{T}$.

{\scriptsize
\begin{subequations}\label{eq:mgder2}
\renewcommand{\theequation}{\theparentequation{}-{}\arabic{equation}}
\begin{align}
&\dfrac{\langle \TT{the}, \lex \TT{=n d}, \epsilon \rangle
    \qquad
    \langle \TT{rat}, \lex \TT{n}, \MT{rat} \rangle
    }
    {\langle \TT{the rat}, \der \TT{d}, \MT{rat}\rangle
    } \: \text{merge-1} \label{eq:mgder2-1} \\
&\dfrac{\langle \TT{eats}, \lex \TT{=n =d c}, \MT{\lambda x.\lambda y.eat(x)(y)} \rangle
    \qquad
    \langle \TT{carrot}, \lex \TT{n}, \MT{carrot} \rangle
    }
    {\langle \TT{eats carrot}, \der \TT{=d c}, \MT{(\lambda x.\lambda y.eat(x)(y))(carrot)} \rangle
    } \: \text{merge-1} \label{eq:mgder2-2} \\
&\dfrac{\langle \TT{eats carrot}, \der \TT{=d c}, \MT{(\lambda x.\lambda y.eat(x)(y))(carrot)} \rangle
    }
    {\langle \TT{eats carrot}, \der \TT{=d c}, \MT{\lambda y.eat(carrot)(y)} \rangle
    } \: \lambda\text{-appl.} \label{eq:mgder2-3} \\
&\dfrac{\langle \TT{eats carrot}, \der \TT{=d c}, \MT{\lambda y.eat(carrot)(y)} \rangle
    \qquad
    \langle \TT{the rat}, \der \TT{d}, \MT{rat}\rangle
    }
    {\langle \TT{the rat eats carrot}, \der \TT{c}, \MT{(\lambda y.eat(carrot)(y))(rat)} \rangle
    } \: \text{merge-2} \label{eq:mgder2-4} \\
&\dfrac{\langle \TT{the rat eats carrot}, \der \TT{c}, \MT{(\lambda y.eat(carrot)(y))(rat)} \rangle
    }
    {\langle \TT{the rat eats carrot}, \der \TT{c}, \MT{eat(carrot)(rat)} \rangle
    } \: \lambda\text{-appl.} \label{eq:mgder2-5}
\end{align}
\end{subequations}
}


\paragraph{\em Forth iteration}
\label{sec:it4}

In the fourth iteration, we suppose that $\mathcal{T}$ utters
\begin{equation}\label{eq:ump4}
    u_4 = \langle \TT{the rats eat cheese}, \MT{eat(cheese)(rats)} \rangle
\end{equation}
that is unified with the previous lexicon $X_3$ through our pattern matching algorithm to yield $X_4$ in \Tab{tab:mg4} in the first place.

\begin{table}[H]
\caption{\label{tab:mg4} Learned minimalist lexicon $X_4$ at time $t = 4$.}
\[
\begin{array}{ll}
    \langle \TT{the}, \lex \TT{=n d}, \epsilon \rangle &
    \langle \TT{mouse}, \lex \TT{n}, \MT{mouse} \rangle \\
    \langle \TT{\underline{rat}}, \lex \TT{n}, \MT{\underline{rat}} \rangle &
    \langle \TT{\underline{rat}s}, \lex \TT{n}, \MT{\underline{rat}s} \rangle \\
    \langle \TT{cheese}, \lex \TT{n}, \MT{cheese} \rangle &
    \langle \TT{carrot}, \lex \TT{n}, \MT{carrot} \rangle \\
    \langle \TT{\underline{eat}s}, \lex \TT{=n =d c}, \MT{\lambda x.\lambda y.eat(x)(y)} \rangle &
    \langle \TT{\underline{eat}}, \lex \TT{=n =d c}, \MT{\lambda x.\lambda y.eat(x)(y)} \rangle
\end{array}
\]
\end{table}

Underlined are again common strings in exponents \index{exponent} or semantics that could entail further revisions of the MG lexicon.

Next, let us assume that $\mathcal{L}$ would express the meaning $\MT{eat(carrot)(rats)}$. It could then attempt the following derivation \eqref{eq:mgder3}.

{\scriptsize
\begin{subequations}\label{eq:mgder3}
\renewcommand{\theequation}{\theparentequation{}-{}\arabic{equation}}
\begin{align}
&\dfrac{\langle \TT{the}, \lex \TT{=n d}, \epsilon \rangle
    \qquad
    \langle \TT{rats}, \lex \TT{n}, \MT{rats} \rangle
    }
    {\langle \TT{the rats}, \der \TT{d}, \MT{rats}\rangle
    } \: \text{merge-1} \label{eq:mgder3-1} \\
&\dfrac{\langle \TT{eats}, \lex \TT{=n =d c}, \MT{\lambda x.\lambda y.eat(x)(y)} \rangle
    \qquad
    \langle \TT{carrot}, \lex \TT{n}, \MT{carrot} \rangle
    }
    {\langle \TT{eats carrot}, \der \TT{=d c}, \MT{(\lambda x.\lambda y.eat(x)(y))(carrot)} \rangle
    } \: \text{merge-1} \label{eq:mgder3-2} \\
&\dfrac{\langle \TT{eats carrot}, \der \TT{=d c}, \MT{(\lambda x.\lambda y.eat(x)(y))(carrot)} \rangle
    }
    {\langle \TT{eats carrot}, \der \TT{=d c}, \MT{\lambda y.eat(carrot)(y)} \rangle
    } \: \lambda\text{-appl.} \label{eq:mgder3-3} \\
&\dfrac{\langle \TT{eats carrot}, \der \TT{=d c}, \MT{\lambda y.eat(carrot)(y)} \rangle
    \qquad
    \langle \TT{the rats}, \der \TT{d}, \MT{rats}\rangle
    }
    {\langle \TT{the rats eats carrot}, \der \TT{c}, \MT{(\lambda y.eat(carrot)(y))(rats)} \rangle
    } \: \text{merge-2} \label{eq:mgder3-4} \\
&\dfrac{\langle \TT{the rats eats carrot}, \der \TT{c}, \MT{(\lambda y.eat(carrot)(y))(rats)} \rangle
    }
    {\langle \TT{the rats eats carrot}, \der \TT{c}, \MT{eat(carrot)(rats)} \rangle
    } \: \lambda\text{-appl.} \label{eq:mgder3-5}
\end{align}
\end{subequations}
}

However, uttering \TT{the rats eats carrot} will probably be rejected by the teacher $\mathcal{T}$ because of the grammatical number agreement error, thus causing punishment by $\mathcal{T}$. As a consequence, $\mathcal{L}$ has to find a suitable revision of its lexicon $X_4$ that is guided by the underlined matches in \Tab{tab:mg4}.

To this aim, the agent first modifies $X_4$ as given in \Tab{tab:mg5}.

\begin{table}[H]
\caption{\label{tab:mg5} Revised minimalist lexicon $X_{41}$.}
\[
\begin{array}{ll}
    \langle \TT{the}, \lex \TT{=num d}, \epsilon \rangle &
    \langle \TT{mouse}, \lex \TT{n -a}, \MT{mouse} \rangle \\
    \langle \TT{rat}, \lex \TT{n -a}, \MT{rat} \rangle &
    \langle \epsilon, \lex \TT{=n +a num}, \epsilon \rangle \\
    \langle \TT{-s}, \lex \TT{=n +a num}, \epsilon \rangle &
    \langle \TT{cheese}, \lex \TT{n}, \MT{cheese} \rangle \\
    \langle \TT{carrot}, \lex \TT{n}, \MT{carrot} \rangle &
    \langle \TT{\underline{eat}s}, \lex \TT{=n =d c}, \MT{\lambda x.\lambda y.eat(x)(y)} \rangle \\
    \langle \TT{\underline{eat}}, \lex \TT{=n =d c}, \MT{\lambda x.\lambda y.eat(x)(y)} \rangle
\end{array}
\]
\end{table}

In \Tab{tab:mg5} the entries for \TT{mouse} and \TT{rat} have been updated by a number licensee \TT{-a} (for \emph{Anzahl}). Moreover, the entry for \TT{the} now selects a number type \TT{=num} instead of a noun. Even more crucially, two novel entries of number type \TT{num} have been added: a phonetically empty item $\langle \epsilon, \lex \TT{=n +a num}, \epsilon \rangle$ selecting a noun \TT{=n} and licensing number movement \TT{+a}, and an item for the plural suffix $\langle \TT{-s}, \lex \TT{=n +a num}, \epsilon \rangle$ with the same feature sequence.

Upon the latter revision, the agent may successfully derive \TT{rat}, \TT{rats}, and \TT{mouse}, but also \TT{mouses}, which will be rejected by the teacher. In order to avoid punishment, the learner had to wait for the well-formed item \TT{mice} once to be uttered by $\mathcal{T}$. Yet, the current evidence prevents the agent from correctly segmenting \TT{eats}, because our shallow semantic model does not sufficiently constrain any further pattern matching. This could possibly be remedied in case of sophisticated numeral and temporal semantic models. At the end of the day, we would expect something alike the hand-crafted lexicon \Tab{tab:mg0} from \Sec{sec:mg}. For now, however, we leave this important problem for future research.


\section{Discussion}
\label{sec:disc}

In this contribution we have outlined an algorithm for effectively learning the syntax and semantics of English declarative sentences. Such sentences are presented to a cognitive agent by a teacher in form of utterance meaning pairs (UMP) \index{utterance meaning pair (UMP)} where the meanings are encoded as formulas of first order predicate logic. This representation allows for the application \index{lambda application} of compositional semantics via lambda calculus \cite{Church36}\index{lambda calculus}. For the description of syntactic categories we use Stabler's minimalist grammar \cite{Stabler97, StablerKeenan03},\index{grammar| minimalist grammar}  (MG) a powerful computational implementation of Chomsky's recent Minimalist Program for generative linguistics \cite{Chomsky95}. Despite the controversy between Chomsky and Skinner \cite{Chomsky59}, we exploit reinforcement learning \cite{Skinner15, SuttonBarto18}\index{reinforcement learning} as training paradigm. Since MG codifies universal linguistic competence through the five inference rules (\ref{merge-1} -- \ref{move-2}), thereby separating innate linguistic knowledge from the contingently acquired lexicon, our approach could potentially unify generative grammar \index{grammar | generative} and reinforcement learning, \index{reinforcement learning} hence resolving the abovementioned dispute.

Minimalist grammar \index{grammar | minimalist grammar} can be learned from linguistic dependency structures \cite{KobeleCollierEA02, StablerEA03, BostonHaleKuhlmann10, KleinManning04} by positive examples, which is supported by psycholinguistic findings on early human language acquisition \cite{Ellis06, Pinker95, Tomasello06}\index{language acquisition}. However, as Pinker \cite{Pinker95} has emphasized, learning through positive examples alone, could lead to undesired overgeneralization. Therefore, reinforcement learning \index{reinforcement learning} that might play a role in children language acquisition \index{language acquisition} as well \cite{Moerk83, SundbergEA96}, could effectively avoid such problems. The required dependency structures are directly provided by the semantics in the training UMPs. \index{utterance meaning pair (UMP)} Thus, our approach is explicitly semantically driven, in contrast to the algorithm in \cite{KleinManning04} that regards dependencies as latent variables for EM training.

As a proof-of-concept we suggested an algorithm for simple English declarative sentences. We also have evidence that it works for German and French as well and hopefully for other languages also. Our approach will open up an entirely new avenue for the further development of speech-controlled cognitive user interfaces \cite{BaranyiCsapoSallai15, TschopeEA18, Young10}\index{user interface}.



\end{document}